\documentclass{article}

\usepackage[final]{neurips_2019}

\usepackage[utf8]{inputenc} 
\usepackage[T1]{fontenc}    
\usepackage{hyperref}       
\usepackage{url}            
\usepackage{booktabs}       
\usepackage{amsfonts}       
\usepackage{nicefrac}       
\usepackage{microtype}      
\title{Methodological Blind Spots in Machine Learning Fairness: Lessons from the Philosophy of Science and Computer Science}

\author{%
  Samuel Deng\\
  Department of Computer Science\\
  Columbia University\\
  New York, NY 10027 \\
  \texttt{sd3013@columbia.edu} \\
  \And
  Achille Varzi\\
  Department of Philosophy \\
  Columbia University \\
  New York, NY 10027 \\
  \texttt{achille.varzi@columbia.edu} \\
}

\begin{document}

\maketitle

\begin{abstract}
In the ML fairness literature, there have been few investigations through the viewpoint of philosophy, a lens that encourages the critical evaluation of basic assumptions. The purpose of this paper is to use three ideas from the philosophy of science and computer science to tease out blind spots in the assumptions that underlie ML fairness: abstraction, induction, and measurement. Through this investigation, we hope to warn of these methodological blind spots and encourage further interdisciplinary investigation in fair-ML through the framework of philosophy.
\end{abstract}

\section{Introduction}
Though the original intent of incorporating machine learning (ML) in human decision-making was to remove human bias, domain applications from criminal justice to hiring have shown that ML might actually exacerbate bias (\cite{hardtbarocas}). This spurred “fair machine learning,” the paradigm of creating ML systems with these biases in mind. In the recent fair ML literature, various mathematical definitions of fairness abound. However, despite being internally consistent through the math, each definition assumes a different philosophical view on fairness (\cite{binns}). For instance, should algorithms give every group the same probability for certain outputs (\cite{statparity})? Or should they instead strive to reduce negative impacts on more disadvantaged groups (\cite{equalodds})? These questions remind us of much older (unsolved) thought in moral philosophy on what “fairness” means in the first place. Thus, it is clear that committing to a fairness definition necessarily forces us to commit to a strong normative view on a definition of fairness. This suggests an alternate philosophical approach – focus on the assumptions and methodology behind ML instead of the existing "solutions." \par

In this paper, we turn our eyes to the methodological assumptions behind ML through the lens of philosophy. We find three methodological "blind spots" through philosophy: abstraction, induction, and measurement. By reframing the investigation as such, we adopt a more holistic approach to fairness that keeps in mind necessary human and societal actors. \par

\section{Fairness as a Context-Sensitive Value}
Before exploring the methodological blind spots in fair-ML, we first make an assumption on the nature of fairness as a value. This initial assumption stems from the intuition that a “fair” ML system must take into account its context and environment (\cite{selbst}). A "value" is defined in the theory of value in moral philosophy, which is broadly concerned with “which things are good or bad and how good or bad are they” (\cite{hirose}). A value $X$ should conform to questioning how $X$ something is. For the value of justice, we might ask, “How \textit{just} is this law?” For our purposes, we define a \textit{context-sensitive value}, a value that yields different evaluations depending on context. Take the value of usefulness. When we say something is useful, it seems intuitive that how useful that thing is depends on the surrounding context. For instance, suppose we compare the usefulness of a chair in a bare room to the last open chair at a table during dinner. In the latter context, the chair is clearly more useful, though both contexts concern the value usefulness. \par

In fair-ML, we argue that fairness is one of these \textit{context-sensitive values} because of ML’s main goal – prediction of phenomena via mathematical formalization on large amounts of data. The “contexts” when implementing fair ML include, but are not limited to: the societal context, the application domain, the data acted upon, the stakeholders, and the algorithms involved. Because the “context” here might be more difficult to parse than in our trivial chair example, we provide an example. \par 

Suppose an algorithm with the fairness definition of Equalized Odds, which equalizes false positive rates between two groups. Then, suppose two application contexts: hiring and criminal recidivism. For hiring, a false-positive (someone unfit for the job is employed) is less harmful than a false-negative (someone fit for the job is denied). An unfit, accepted applicant might just get fired, while there is no “second chance” for a fit but unaccepted applicant. However, in criminal recidivism, equalized false-positives might keep more of a minority population mistakenly in jail, propagating an existing injustice. Instead, we might use a different fairness definition altogether for recidivism. Our single mathematical formalization fails to generalize to different social contexts (\cite{selbst}). Fairness might also be sensitive to the context of time. Our notions of what is fair drastically evolve over time periods or societal eras. Thus, fairness in the specific domain of fair-ML is a \textit{context-sensitive value}. \par

\section{Blind Spot 1: Abstraction}

The first blind spot fair ML has with respect to context-sensitive fairness is abstraction, an integral tenet in the philosophy of computer science. In ML, mathematical abstractions allow models to become black-boxes. Here, we focus on two aspects of abstraction: specification and modularity.

First, the notion of specification in fair-ML is too narrow for context-sensitive fairness. Any algorithm involves a specification -- a description of its inputs and outputs. A simple specification might be "input $x$ and output its square, $x^2$." Well-defined specifications allow “black box” abstractions to build into complex systems. In fair-ML, we might have the standard specification of optimizing a loss function with the additional specification of a fairness constraint. We argue that this is not enough -- fairness requires a human, normative specification, with respect to the semantics of fairness in a context. A mathematical formulation of fairness provides only the syntax. For instance, some fairness constraint might "equalize false positive rates," but true fairness calls for a normative specification: "...assuming equal false positive rates are acceptable due to the moral motivation for long-run diversity between groups A and B." Without this, ML algorithms are merely "symbol-crunching," as Searle argues in his famous Chinese Room Argument -- while machines might use syntactic rules to manipulate symbols, they cannot grasp semantic values (\cite{searle}). To get past "just doing the math” of fairness, human domain experts in the loop are necessary to provide concrete normative specifications. \par

Second, abstraction's end goal of modularity (allowing for the "plug and play" necessary for complex computer systems) runs counter to context-sensitive fairness. In ML, problems are characterized by their nature of learning task, and a specific set of established methods (e.g. SVM, nearest neighbors) are implemented as single line imports in ML libraries (e.g. scikit-learn, TensorFlow). Some fair ML applications even utilize a "fair wrapper" to encourage a modular approach for fairness (\cite{selbst}). However, modularity and context-sensitive "fairness" are mutually exclusive. Given an ML system with satisfactory fairness constraints in a specific context, we cannot ship it to another context and expect it to still be "fair." For instance, suppose a fair system that equalizes false positive rates for hiring by fulfilling all normative assumptions for fairness in hiring. We cannot simply "export" it to the domain of criminal justice, where the normative value of a false positive is far different. 

\section{Blind Spot 2: Induction}
The second blind spot is induction, a central problem in the philosophy of science. Induction is the principle that future unexperienced instances of something resemble those of which we have had experience. Hume posed the unsolved "problem of induction" -- inductive patterns are no more than "habits of the mind," and there is no necessary connection between observations and unseen events (\cite{hume}). This led to Nelson Goodman's "new riddle of induction," (\cite{goodman}) which we use to motivate how the problem of induction affects ML. Goodman proposes the thought experiment: suppose that, up to time $t,$ we have observed many green emeralds and no emeralds other than green ones. He makes a first hypothesis: "All emeralds are green." Goodman then introduces the predicate "grue": an object is "grue" if, observed before a time $t,$ it is green, but, after $t,$ it is blue. Note that the observations also fulfill, "emerald $x$ before time $t$ is "grue." Thus, Goodman makes another hypothesis: "All emeralds are grue." Goodman’s two hypotheses are equally supported by the data, given our new predicate, but “All emeralds are green" expects the next emerald to be green, while "All emeralds are grue" expects the next to be blue. Therein lies the problem -- for the same set of data, how do we distinguish the good inductions from the bad inductions? Intuitively, "all emeralds are green" is a better induction, but with just the data and hypotheses at hand, there is nothing to necessitate this. \par

Learning theory (the mathematical theory behind ML) answers Goodman's problem with its cornerstone assumption: Occam's Razor, the principle that one should choose the simplest explanation of a phenomenon compatible with one's experiences. This allows for "All emeralds are green." Then, because learning theory is the study of "computational strategies for converging to the truth," (\cite{harizanov}) the Occam's Razor assumption conflates the most truth-conducive hypothesis with the simplest. This embeds a crucial philosophical assumption deep into ML's theoretical underpinnings. Indeed, attempts in the growing field of ML interpretability have begun to grapple with this blindspot (\cite{1602.04938}, \cite{1702.08608}). \par

In fair-ML, Occam's Razor conflicts with context-sensitive fairness in three ways. First, ML systems generalize the "simplest explanation" of historical trends in data that may have a normatively charged context. For instance, suppose ML for hiring in a workplace using past resume data, where the past twenty years have been wrought with gender bias against women. Following Occam's Razor, our algorithm hypothesizes the simplest explanation that "women perform worse," because the training data's most expressive signal is that women do not get hired. This clearly conflicts with context-sensitive fairness if our context is a society that aims to progress from past injustices prevalent in all previously observed instances. Second, learning-theoretic induction might fail us when ML picks up on nonsensical trends in data. Take a recent ML paper that created a highly accurate classifier for criminals using only portraits (\cite{wu}). A rebuttal found that this algorithm simply classified criminals with high accuracy based on whether they frowned in their portraits. Here, Occam's Razor allows potentially unfair inductive hypotheses through highly expressive and "simple," yet nonsensical trends in the data. Third, Occam’s Razor has no robustness against feedback loops. In a well-known example, predictive policing might use arrest records to predict the location of new crimes and determine police deployment. But this causes increased surveillance and, with it, arrest records on those very neighborhoods (\cite{barocas-hardt-narayanan}). Here, feedback loops cause an algorithm to pick up on the simplest, feedback-amplified hypothesis from the data, but the more complex hypothesis (that there is a feedback loop), remains undetected until human intervention. \par

\section{Blind Spot 3: Measurement}
In the philosophy of science and epistemology, measurement is rich grounds for philosophical inquiry. Measured data is central to ML before any algorithm is even implemented, so the crucial question is: how do we measure this data? The dominant theory of measurement is the representational theory of measurement (RTM) (\cite{krantz}). RTM formalizes measurement as the construction of mappings from empirical structures to formal structures. This mapping is known as a scale. For instance, we might try to measure wooden rods $A$ and $B$. To construct a scale, we might define that rod $A$ is longer than rod $B$ if, when one of the ends of each rod are aligned, the other end of rod $A$ extends past rod $B$. This is the empirical relationship of "longer." By assigning real numbers to the rods and the symbol $>$ to the relationship "longer" we may write $A > B$ as the formal relationship. Here, our empirical assumption in the mapping is a relatively harmless procedure to distinguish the relationship of "longer" – namely, putting the rods flush to each other – but it is an assumption, nonetheless. In measurement, these assumptions are necessary. We show that these assumptions can conflict with context-sensitive fairness in more complex examples. \par

With RTM in mind, we first argue that not all formal scales can be liberally “statisticized." Because a scale is a mapping from the empirical to the formal, different empirical relationships admit different formal scales (e.g. nominal, ordinal, interval). However, some scales do not admit of common statistical techniques (\cite{hardtbarocas}). For instance, restaurant reviews may be measured on an ordinal scale from 1 to 5. In an ordinal scale, a 2 is better than a 1, and so forth, but just how much better is an undefined quantity. We only have an ordering of objects. Because of this, taking basic statistical operations such as the mean or standard deviation is fundamentally a category error (\cite{hardtbarocas}), as they work with more than just ordering. Despite this, data in many ML applications might be on an ordinal scale (e.g. restaurant reviews, movie ratings, patient surveys). By imposing a scale on empirical phenomena, we risk assumptions that are incompatible with ML, and, by applying invalid operations on these biased assumptions, fair-ML systems face a serious methodological issue. \par

Second, on the empirical side of measurement, data features and labels often come from measuring elusive empirical concepts. In fair-ML, measuring data features oftentimes means measuring features related to humans. For instance, an ML system for hiring might want to measure empirical features such as intelligence and communicability. Possible measures of such traits might be an IQ test or a written sample. But who decides the measurement, and how do we know if an IQ test is a good measurement of intelligence? And how do we score a writing sample without introducing reader bias? In many cases, the data features of our inputs already incorporate context dependent assumptions. As Moritz Hardt put it in NIPS 2017, "every feature is a model" 
(\cite{hardtbarocas}). That is, features incorporate important normative and subjective context-dependent assumptions about measurements, but all features are treated as numerical truth by ML systems. This also applies to the labels in supervised learning, as they are often constructs created only for the purpose at hand. For instance, when building an ML system for job performance, we might only have the proxy of historical performance reviews (\cite{barocas-hardt-narayanan}). However, historical performance might admit all sorts of biases, such as the culture of past managers or different teams in the workplace. Thus, in both features and labels, context is crucial to understanding the measurements involved, as training data is given as "ground truth" for ML models. However, there is no current approach of encoding contexts \textit{themselves} into ML features or labels. Nascent empirical attempts to survey \textit{actual} human judgment on the fairness of training features have shown that human evaluation reveals numerous concerns that the normative attempts at fairness in the literature have not yet grappled with (\cite{1802.09548}).  Without some incorporation of context, these applications fail to satisfy context-sensitive fairness. \par

\section{Conclusion and Suggestions}
We have shown three methodological objections to fair-ML from philosophy: abstraction, induction, and measurement. These objections stem from the assumption that fairness is a context-sensitive value, and, as such, avoids capture through methods such as simply specifying a “one-size-fits-all” definition of fairness (abstraction) or assuming we have well-measured data (measurement). We conclude with three broad suggestions for ways in which ML researchers, philosophers, and domain experts might approach fair-ML in a way that embraces the context-sensitivity of fairness and begin to assuage these deep-rooted philosophical issues. \par

First, \textit{abstraction in fair-ML should begin the iterative process with normative assumptions.} Fair ML systems cannot iterate just on the technical portion of the process, driving up accuracy until we notice some societally unfavorable result. Instead of checking for normative assumptions and fairness at the end of the pipeline, domain experts and ML practitioners should incorporate context-dependent normative assumptions into the initial specifications of the system. This interdisciplinary approach integrates human domain experts who truly understand the context of the system as necessary, ensuring that values are not lost in the process before abstraction takes place.\par

Second, \textit{when fairness is involved in induction, the simplest inductive patterns are not always the fairest.} By better understanding that the simplest induction is not always the fairest induction when we review the context of each case in fair-ML, ML practitioners and domain experts alike must engage in dialogue about possible alternative hypotheses on the data. This requires an open-mindedness to the wide universe of “not-so-simple” hypotheses (historical trends, feedback loops) and cases where (as proposed in Section 4) emeralds are actually “grue” instead of “green.” \par

Third, \textit{take a closer look at the measurement of data, the representation of those measurements, and question “ground truth.”} A better understanding of measurement as a subjective mapping from the empirical to the formal might inspire deeper inquisition into the measurement of certain labels or features in data. In the domain of fair-ML, this is especially important, as, oftentimes, the traits we must measure are human or social traits that elude simple formal schemes. Again, open dialogue is needed between domain experts, ML practitioners, and, very importantly, social scientists (who have better understanding of how human qualities can be quantified) to take a closer look at the ways we measure our data and come up with ways to improve.\par

While technical and concrete solutions await, the advantage of philosophy is that it allows us to revisit our assumptions, see our blind spots, and recognize when we might be “losing the forest for the trees.” We can only hope that, moving forward, philosophy is a more widely used tool in ML to tease out and question its most basic assumptions. \par

\bibliography{references}

\end{document}